\documentclass{aa}

\usepackage{txfonts}
\usepackage{epsfig}

\begin{document}

\title{Numerical self-consistent stellar models of thin disks}

\author{Maximiliano Ujevic\thanks{e-mail: mujevic@ime.unicamp.br} \and 
Patricio S. Letelier\thanks{e-mail: letelier@ime.unicamp.br}}

\institute{Departamento de Matem\'atica Aplicada, Instituto de
Matem\'atica, Estat\'{\i}stica e Computa\c{c}\~ao Cient\'{\i}fica,
Universidade Estadual de Campinas, 13081-970, Campinas, SP, Brasil}

\date{}

\abstract{We find a numerical self-consistent stellar model by finding
the distribution function of a thin disk that satisfies simultaneously
the Fokker-Planck and Poisson equations. The solution of the
Fokker-Planck equation is found by a direct numerical solver using
finite differences and a variation of Stone's method. The collision term
in the Fokker-Planck equation is found using the local approximation and
the Rosenbluth potentials. The resulting diffusion coefficients are
explicitly evaluated using a Maxwellian distribution for the field
stars. As a paradigmatic example, we apply the numerical formalism to
find the distribution function of a Kuzmin-Toomre thin disk. This
example is studied in some detail showing that the method applies to a
large family of actual galaxies.

}

\titlerunning{Numerical self-consistent stellar models of thin disks}
\authorrunning{M. Ujevic \& P.S. Letelier}

\maketitle

\section{Introduction}

Few exact analytic stationary axisymmetric solutions for the
collisionless Boltzmann equation are known in astrophysics. Most of the
solutions correspond to spherical systems such as the polytropes and
Plummer's models (Schuster \cite{sch}; Plummer \cite{plu}, Chandrasekhar
\cite{cha}), and lowered isothermal models known as King models (Michie
\cite{mic}; Michie \& Bodenheimer \cite{mic:bod}; King \cite{kin}) which
describe well globular cluster data. Also, we can find self-consistent
spherical models by using Eddington's formula (Eddington \cite{edd}). 
Furthermore, at least half of spiral galaxies and most of the elliptical
galaxies appear to be triaxial stellar systems. Models for triaxial
systems have been constructed numerically, usually end products of
$N$-body collapse simulations (Aarseth \& Binney \cite{aar:bin}; Miller
\& Smith \cite{mil:smi1},b; van Albada \cite{alb}; Wilkinson \& James
\cite{wil:jam}) or using catalogs of numerically integrated orbits
(Schwarzschild \cite{schw}; Statler \cite{sta}; Teuben \cite{teu}); and
analytically for special cases (Freeman \cite{fre}; Vandervoort
\cite{van}). Moreover, other self-consistent triaxial models that
satisfy the St\"ackel potentials were studied (de Zeeuw \cite{zee}; de
Zeeuw et al. \cite{zee:hun}; Hunter \& de Zeeuw \cite{hun:zee}).
Recently, a general solution of the Jeans equation for triaxial systems
has been found (van de Ven et al. \cite{ven:hun}). But, in many
astrophysical objects light is emitted by a thin stellar disk, so planar
galaxies are of more than academic interest. The most important exact
analytical solutions of the collisionless Boltzmann equation for thin
disks are the ones due to Mestel (\cite{mes}) and Kalnajs (\cite{kal}).
Also, Miyamoto (\cite{miy}) found an approximated analytical solution
for the Toomre disk using a series method, and Ng (\cite{ng}) found a
self-consistent model for a thin stellar system, from a given
distribution function, performing numerical calculations. The importance
of having stationary solutions, for stellar objects or other systems, of
the Boltzmann or Fokker-Planck equations is that they are a starting
point for doing dynamical evolutions, and to study, analytically or
numerically, the stability of the system by performing different kinds
of perturbations. These studies have been done over the years by many
authors, as examples we can mention (Cohn \cite{coh}; Watanabe et al.
\cite{wat:ina}; Nishida et al. \cite{nis:wat}; Yepes \&
Dom\'{\i}nguez-Tenreiro \cite{yep:dom}; Takahashi \cite{tak}; Theuns
\cite{the}; Takahashi et al. \cite{tak:lee}; Joshi et al.
\cite{jos:nav}; Ashurov \cite{ash}).  In most of the thin disk models,
we only have as information their Newtonian gravitational potentials and
surface mass density, knowing its distribution function we can obtain
the thermodynamic variables allowing a full understanding of the physics
of each individual model. Thus, the solution of the Fokker-Planck
equation for each particular disk is needed to determine its main
physical properties. Despite of the importance of the Kalnajs and Mestel
thin disks, we have to keep in mind that they do not model a realistic
physical situation because they exhibit odd behaviors within the disk,
i.e. in the Kalnajs disks, the distribution function $f$ goes to
infinity for certain values of the energy density and angular momentum;
and in the Mestel disks, the gravitational potential $\Phi$ diverge at
the center of the disk.

The Fokker-Planck equation is also known as the Fokker-Planck
approximation because it truncates the BBGKY (Bogoliubov, Born, Green,
Kirkwood, and Yvon) hierarchy of kinetic equations, at its lowest order,
by assuming that correlation between particles only plays a role as a
sequence of uncorrelated two-body encounters. It is worth noticing that
the only ``approximation'' made in the Fokker-Planck equation comes from
the model adopted for collisions and, in fact, the Fokker-Planck
equation with a general collision term can be derived from first
principles and no {\it ad hoc} suppositions are needed. The main goal of
our work is to find and study a self-consistent model of a stellar thin
disk, i.e. a distribution that satisfies simultaneously the
Fokker-Planck and the Poisson equation, the meaning of simultaneously
will become clear later in the manuscript. In writing the Fokker-Planck
equation, we considered in the collision term the local approximation
and used the diffusion coefficients found by Rosenbluth et al.
(\cite{ros:mac}).

The computational problems in solving the Fokker-Planck equation without
using any statistical (Monte Carlo method) and average (moment
equations) are briefly discussed in this paper. The Fokker-Planck
equation is solved by a direct numerical solver using finite differences
and a variation of Stone's method (Stone \cite{sto}; Ujevic \& Letelier
\cite{uje:let2}) to include mixed derivatives. This allows high
resolution solutions with lower computational cost (Ujevic \& Letelier
\cite{uje:let2}). Due to nature of the problem, either analytical or
numerical solutions are very difficult to obtain. In this respect, it is
illustrative to cite the words of Binney and Tremaine (\cite{bin:tre}
page 245): {\it Finding the particular function of three variables that
describes any given galaxy is no simple matter. In fact, this task has
proved so daunting that only in the last few years, three-quarters of a
century after Jeans's} (Jeans \cite{jea}) {\it paper posed the problem,
has the serious quest for the distribution function of even our own
Galaxy got underway.} As far as we know, there are no analytical or
numerical solution of the Fokker-Planck equation for thin disks.

The article is organized as follows. In Section II we present the
Fokker-Planck equation for a thin disk with the collision term given by
the local approximation and the diffusion coefficients found by
Rosenbluth et al. (\cite{ros:mac}). In Section III, we explain briefly
the computational codes used to solve the Fokker-Planck equation. In
Section IV, we present the properties of the Kuzmin-Toomre thin disk
used as an example for the numerical method. The numerical results for
the velocity and spatial distribution functions for galaxies based in
this model are showed. Finally, in Section V, we summarized our results.

\section{Fokker-Planck equation for a thin disk}

The general problem for the evolution of a system made with stars of
mass $m$ using the Fokker-Planck equation consists in the determination
of a distribution function $f$ and the gravitational potential $\Phi$
that solve the Fokker-Planck and Poisson equations, say

\begin{eqnarray} 
&&\frac{\partial f}{\partial t} + {\bf v} \cdot \nabla f - \nabla \Phi
\cdot \nabla_v f = \Gamma[f], \label{fokker} \\ 
&&\nabla^2 \Phi = 4\pi G m \int f d^3v,
\label{poisson}
\end{eqnarray}

\noindent where ${\bf v}$ represents the velocity of the stars, $\nabla$
is the usual spatial gradient, $\nabla_v$ is the velocity gradient
(derivations are done with respect to the velocities), $G$ is the
gravitational constant and the symbol $\Gamma[f]$ represents the
Fokker-Planck collision term.  The solution of the Fokker-Planck and
Poisson equations is found from an iterative procedure in an short time
interval with respect to the time scale in which $\Phi$ varies. The
method of evolution has two main steps: (i) the distribution function is
first advanced in time using equation (\ref{fokker}) with the potential
held fixed, and (ii) the potential is then adjusted so that the Poisson
equation is again satisfied. Some systems, as a final state of the
evolution, may reach a stationary regime. Then, the temporal part in
(\ref{fokker}) is equal zero and the distribution function is time
independent. Hence, in this case, the distribution function must satisfy
simultaneously the stationary Fokker-Planck equation

\begin{equation} 
{\bf v} \cdot \nabla f - \nabla \Phi \cdot \nabla_v f = \Gamma[f], 
\label{sfokker} 
\end{equation}

\noindent together with equation (\ref{poisson}). Numerically we have
that, in a non-stationary regime, a gravitational potential at time $t$
creates through the Fokker-Planck equation a distribution function at a
later time $t+\Delta t$. In our case, a stationary regime, we want to
find a gravitational potential and a distribution function at the same
time $t$. This is what we mean when we say that the solution ($\Phi,f$)
satisfies simultaneously the Fokker-Planck and Poisson equation. Note
that we can start with a given gravitational potential at time $t$ and
make the evolution of the system by first finding its distribution
function at a time $t+\Delta t$, and later, the new gravitational
potential, via equation (\ref{poisson}), at this new time $t+\Delta t$,
and hope that the successive iterations of this procedure converge. In
the case of convergence, we obtain also a stationary solution but this
solution might not have any physical significance, i.e. the final
gravitational potential together with the distribution function do not
model any astrophysical system. For that reason, it is important to find
the simultaneous stationary solution of the Fokker-Planck equation for a
given gravitational potential that can describe a real system. The
procedure to find this solution is explained in the next section. Our
main goal is to obtain a physical consistent model by solving the set of
equations (\ref{poisson},\ref{sfokker}) without using any
approximations, like the ones found in the orbit averaged approximation,
Monte Carlo method, or the Jeans equations (Binney \& Tremaine
\cite{bin:tre}), i.e. a direct numerical solution.

To find the stationary Fokker-Planck equation valid for thin disks, the
non-stationary case is straightforward, we start from the general case
in three dimensions and later perform the projection to the thin disk
plane.  The collision term in the local approximation has the form

\begin{eqnarray}
&&\Gamma[f]= - \sum_{i=1}^3 \frac{\partial}{\partial v_i} [ f({\bf x,v})
D(\Delta v_i)] \nonumber \\
&&\hspace{1.cm} + \frac{1}{2} \sum_{i,j=1}^3 \frac{\partial^2}{\partial 
v_i \partial v_j} [f({\bf x,v}) D(\Delta v_i \Delta v_j)], 
\label{collision}
\end{eqnarray}

\noindent where the functions $D(\Delta v_i)$ and $D(\Delta v_i \Delta
v_j)$ are known as the diffusion coefficients. These coefficients were
calculated (Rosenbluth et al. \cite{ros:mac}) assuming and inverse
square force for the interaction, and also, that each stellar encounter
involve only a single pair of stars and are independent of all others.
They are simplified if the field stars distribution function is a
Maxwellian distribution (Binney \& Tremaine \cite{bin:tre}). Recall that
the diffusion coefficients are calculated considering a test star of
mass $m$ moving through an infinite homogeneous sea of field stars of
mass $m_a$ who has mean velocity equal to zero. We know from stabilities
analysis that this gravitating system can not be in static equilibrium
and to overcome this problem we invoke the Jeans swindle. In order to
calculate the diffusion coefficients in the next sections, we are going
to set the dispersion of the field stars equal to the typical velocity
of the problem considered.

The diffusion coefficients described above are valid for a general three
dimensional case. To obtain the Fokker-Planck equation for thin disks,
we perform a projection on the $z=0$ and $v_z=0$ planes using Dirac
deltas. Doing the replacement

\begin{eqnarray}
f(x,y,z,v_x,v_y,v_z) \rightarrow f(x,y,z,v_x,v_y,v_z) \delta(z) 
\delta(v_z), \label{delta} 
\end{eqnarray}

\noindent in (\ref{fokker}), and integrating in the $z$ and $v_z$
variables, we arrive at a stationary Fokker-Planck equation in two
dimensions, see Appendix \ref{deltas}. The indexes (i,j) of the
collision term (\ref{collision}) now take only the values 1 and 2. The
important thing about this procedure is that the form of the
Fokker-Planck equation and the diffusion coefficients are preserved but
projected into the $z=0$ and $v_z=0$ planes.

\section{Numerical method}

The solution of the Fokker-Planck equation in a stationary regime is not
an easy task. The problems involve in the solution are both numerical
and physical. Numerical because in the three dimensional case, the
Fokker-Planck equation has six independent variables, three space
coordinates ({\bf x}) and three velocity coordinates ({\bf v}). In the
two dimensional case, the disk case, a simplification exist because the
total number of variables are four. In either cases, in a finite
difference scheme, the large number of grid nodes needed for the
computation of the solution becomes a data storage problem. In a three
dimensional problem, if we divided each of the six variables intervals
of the distribution function in nine parts (ten nodes), we will have a
grid with $10^6$ nodes. This corresponds to a same number of linear
equations to solve. In a simple numerical method we have to store and
solve a matrix with $10^{12}$ elements. For two dimensional disks, the
main matrix will have $10^8$ elements. So, our problem is slightly
simpler than the three dimensional case. With 10 grid nodes per variable
only very simple geometries can be described. Furthermore, the large
number of matrix elements brings us another computational problem, the
slowness of the codes. Most of the elements of the these matrices are
zero and this motivates the search for alternative and faster methods to
solve the linear system using only the non null data. We overcome these
problems performing a discretization of the Fokker-Planck equation using
a finite difference thirteen-point molecule method and then solving the
resulting discretization matrix with a modification of the method
proposed by Stone (\cite{sto}). The original method of Stone was
modified to handle two (four variables) and three (six variables)
dimensions, and the extra diagonals that appear when mixed derivatives
are used (Ujevic \& Letelier \cite{uje:let2}). Also, it can handle
non-stationary problems, and curve boundary conditions with a little
modification of the code. Remember that the collision term in the
Fokker-Planck equation has mixed derivatives. Physical because the
search for a boundary condition is not always clear. In general, the
analytic potentials that we have are only geometrical, i.e they have
been developed without using any observational data (only an image
method solution for a point mass). From this fact, it is not obvious
that these potentials could represent some real galaxies.

The surface mass density of the disk is obtained integrating the right
hand side of (\ref{poisson}), say

\begin{equation}
\Sigma=4 \pi G m \int_{-\infty}^{\infty} f d^3v. \label{mass}
\end{equation}

\noindent Due to the fact that the integral limits of (\ref{mass}) are
from $-\infty$ to $\infty$, we must define first a criteria to establish
the velocity domains for the calculation of the distribution function.
In general, we are going to use the value of the escape velocity
$v_{esc}= \sqrt{2 |\Phi|}$ at the center of the disk to establish the
upper and lower limits of the integration.

When Cartesian coordinates are used, the spatial disk part of the
distribution function will be delimited by an irregular boundary
(circumference). By using cylindrical coordinates we can transform the
irregular domain into a more simple square domain, but in this case we
need more boundary conditions and assumptions for the distribution
function. For example, if we are using a Dirichlet boundary conditions,
we need the value of the distribution function on the side of the square
that corresponds to the coordinate $R=\sqrt{x^2+y^2}=0$ (the center of
the thin disk), a value that is not always known. To avoid these
difficulties, we prefer to use Cartesian coordinates in the spatial
domain and treat the problem of the irregular boundary, the code
presented in (Ujevic \& Letelier \cite{uje:let2}) allows this option. In
compensation, we only need the value of $f$ in the exterior radius of
the disk, i.e. only one expression for the boundary condition.

\begin{figure} 

\epsfig{width=5.cm,height=5.5cm,file=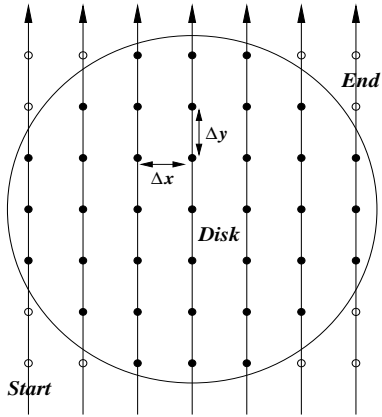}

\caption{Nodes in the spatial domain to be used in the two dimensional
figure of the surface mass density. The full circles represents the nodes
inside the disk used for the calculation. The empty circles are necessary
for maintaining an order in the grid but do not enter into the solution.  
After finding the surface mass density, the spatial part of the solution
of the distribution function can be divided into vectors (represented like
arrows). This procedure help us to present the surface mass density
results in a simple way using a two dimensional figure.}
\label{figgrid3}
\end{figure}

The integration program was tested by integrating exponential functions
of the form exp[$(-v_x^2-v_y^2)/2\sigma^2$] for different values of
$\sigma$. The integral is performed using Romberg's integration scheme
with Richardson style extrapolation (Gerald \& Wheatley \cite{ger:whe}).
The Fokker-Planck code was tested comparing the results obtained with
the modified Stone method with other methods, as for example, the
explicit Euler method and Gauss elimination, obtaining full agreement. 
This was performed for a small number of nodes because for larger grids
the usual methods are not convenient. All these tests assure us that our
codes are working properly. The $LU$ decomposition used in the modified
Stone method is presented in the Appendix \ref{lu}.

The general procedure to find the distribution function of a thin disk
is as follows: first we start with a given stationary disk potential
which is inserted into the Fokker-Planck equation. Then, we find
numerically a distribution function in phase space associated to the
gravitational potential and the boundary condition imposed. Later, we
perform the integration (\ref{mass}) using the recently calculated
distribution function to find the surface mass density of the problem.
This surface mass density has to be equal to the surface mass density of
the given gravitational potential if we want to have stationary
self-consistent solutions between the Fokker-Planck and Poisson
equations. If the surface mass density obtained after the integration of
(\ref{mass}) is not equal to the one given a priori then it is an
indication that the boundary condition used for the distribution
function is not compatible with the problem studied. Our goal is to find
the self-consistent solution of the problem by adjusting the parameters
involve in the gravitational potential, galaxy model and boundary
condition of the Fokker-Planck equation. Note, that in this procedure we
do not solve the Laplace equation. This was done only for didactic
reasons and because there exist in the literature a huge variety of
methods that solve this equation. Actually, to obtain a self-consistent
model, we only need the surface density of the thin disk. With this
information, we find the gravitational potential via Laplace equation,
and later we apply the above procedure.

\begin{figure*}
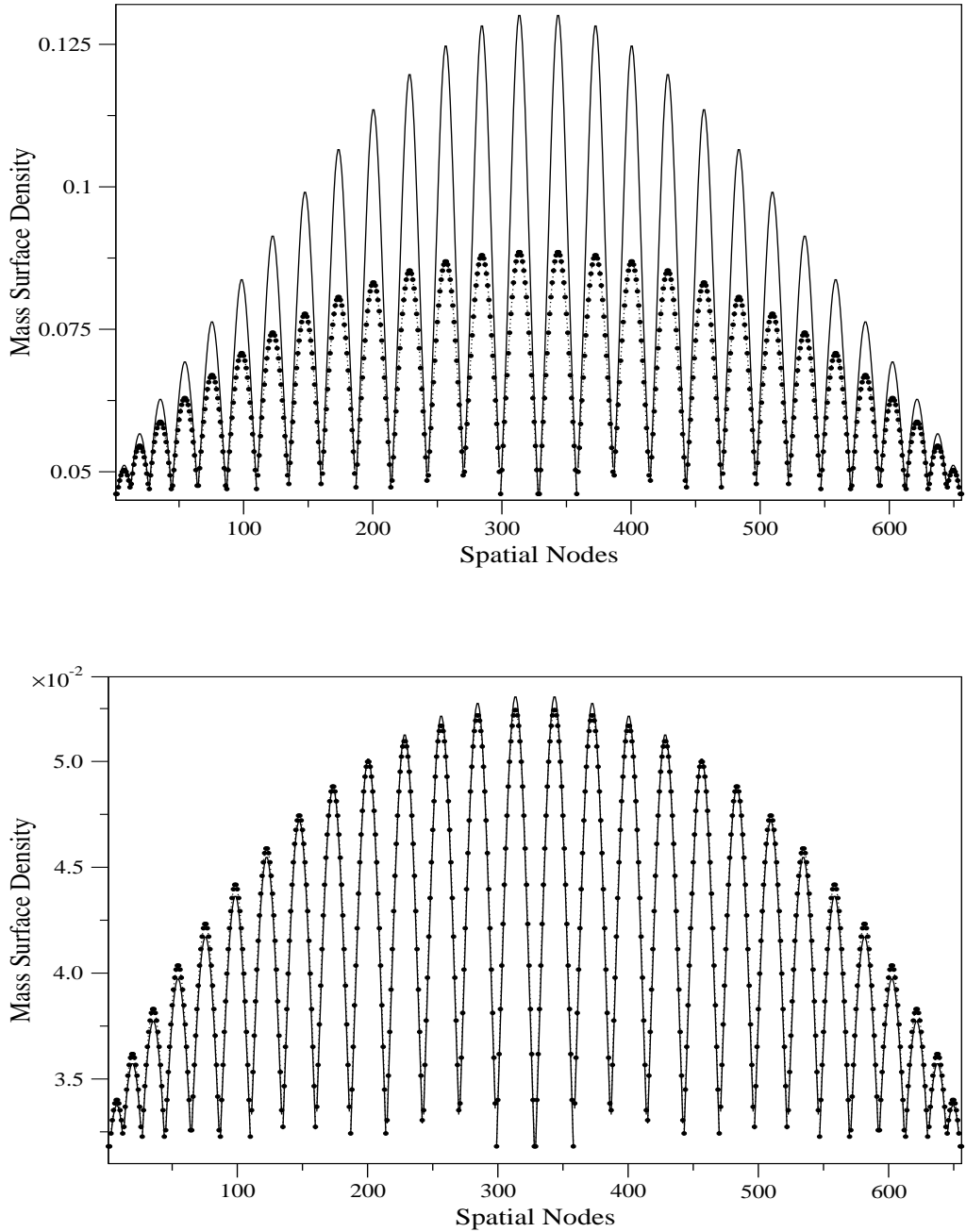


\vspace{1.2cm}
\epsfig{width=13.5cm,height=8cm,file=3176fig2.eps} 

\vspace{1.3cm}
\epsfig{width=13.5cm,height=8cm,file=3176fig3.eps}
\vspace{0.2cm}

\caption{ Top: Two dimensional plot of the numerical solution for the mass
surface density of the Kuzmin-Toomre like disk with cut parameter $a=40$
kpc, and $\sigma_s = 35$ kpc. The solid line is the exact solution
obtained from the mass surface density of the Kuzmin-Toomre disk type. The
dotted line represents the mass surface density obtained after the
integration of the numerical solution, the full circles are the values of
the numerical calculations. The arrows from Fig. \ref{figgrid3} are
plotted one along the others. Each point represents a position $\{x,y\}$
on the surface of the disk. Bottom:  the same of Top but with cut
parameter $a=62.7$ kpc and $\sigma_s = 40$ kpc. Adjusting the values of
the cut parameter $a$ and the spatial dispersion $\sigma_s$ we can
approach the numerical mass surface density profile to the exact
solution.}
\label{density} 
\end{figure*}

Now, we need a boundary condition for $f$. An appropriate boundary
condition can be found if we recall that the velocity distribution in a
star system is determined by two competing processes: relaxation through
stellar encounters that drives the distribution toward a Maxwellian
form, but stars beyond the finite escape velocity continually disappear
from the system. Since escaping stars traverse the system in a small
fraction of a relaxation time, the resultant velocity distribution
should be zero beyond the escape velocity but otherwise nearly
Maxwellian. Also, most of the stars in thin galaxies are centrally
concentrated. So, for a stationary system, these considerations lead us
to use as an appropriate boundary condition for our disk the function

\begin{equation}
f_{Boun} \propto {\rm exp}[(-x^2-y^2)/2 \sigma_s^2] \; {\rm 
exp}[-(v-v_{\rm typ})^2/ 2\sigma_v^2], \label{boundary}
\end{equation}

\noindent where $\sigma_s$ and $\sigma_v$ are the spatial and velocity
dispersion respectively, and $v =\sqrt{v_x^2+v_y^2}$. The velocity
dispersion $\sigma_v$ is a measure of the deviation from the star
typical velocity of the galaxy. The spatial dispersion $\sigma_s$ is a
measure of how centrally concentrated these stars will be. The velocity
part of the boundary condition is centered at the typical velocity of
the system.

\section{Kuzmin-Toomre Galactic Model}

The code used, and the procedure presented in this manuscript, can be
applied, in principle, to find the distribution function for any thin
disk potentials. But, as an example, we are going to used the
Kuzmin-Toomre family of thin disks (Kuzmin \cite{kuz}; Toomre
\cite{too}; Binney \& Tremaine \cite{bin:tre}) which, in cylindrical
coordinates, has the Newtonian gravitational potential,

\begin{equation}
\Phi_{TK}(R,z)=-\frac{GM}{\sqrt{R^2+(a+|z|)^2}}. \label{pottoom}
\end{equation}

\noindent When $z < 0$, $\Phi_{TK}$ coincides with the potential of a
point mass $M$ located at the point $(R,z)=(0,a)$; and when $z >0$,
$\Phi_{TK}$ is exactly the potential of a point mass at $(0,-a)$,
hereafter $a$ is denoted as the cut parameter. The cut parameter
modifies physical properties of the disk, like energy density, pressures
and sound velocity profiles (Vogt \& Letelier \cite{vog:let}). By
applying Gauss's theorem to a flat volume that contains a small portion
of the plane $z=0$, we conclude that the potential (\ref{pottoom}) is
generated by the surface density

\begin{equation}
\Sigma_{TK}(R)=\frac{aM}{2 \pi (R^2+a^2)^{\frac{3}{2}}}. \label{dentoom}
\end{equation}

\noindent The potential (\ref{pottoom}) describes a disk which extends
from $R=0$ to $R=\infty$. In this work we shall consider a disk with
finite radius with the same type of potential.

We start with a galaxy model of radius 40 kpc containing $10^{12}$ stars
of solar masses $M_\odot$. The typical streaming velocity considered is
of order 200 km s$^{-1}$ with velocity dispersion $\sigma_v=40$ km
s$^{-1}$. Hereafter, we set the masses of the field and test stars equal
($m = m_a =$ $M_\odot$). The mass density $\rho$ that appears in the
diffusion coefficients is calculated spreading the total mass of the
system in a thin disk, which radius is given by the model. To solve the
Fokker-Planck equation for the Kuzmin-Toomre disk type potential we use
a total of 810000 nodes, $30^4$, distributed along the computational
domain. All the nodes are necessary to maintain a computational order
that is needed by the code, but actually we do not used all of these
nodes in the calculation because some of them lie outside the thin disk
region, see Fig. \ref{figgrid3}. Recalling that, we have to check the
results of our numerical solution of the distribution function
indirectly, we compare the surface mass density obtained from the
integral (\ref{mass}) with the exact solution (\ref{dentoom}). This will
give us an estimate of how far we are from the exact distribution
function. In Fig.  \ref{figgrid3} we present the spatial part of a disk
with a given number of nodes used for the computation. 

\begin{figure}
\epsfig{width=8cm,height=6cm,file=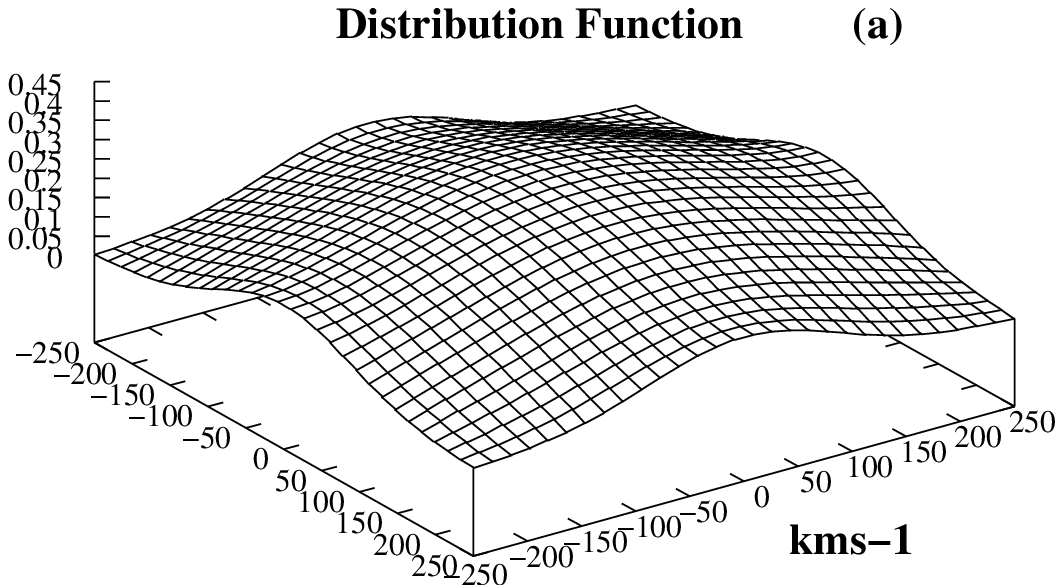}
 
\epsfig{width=8cm,height=6cm,file=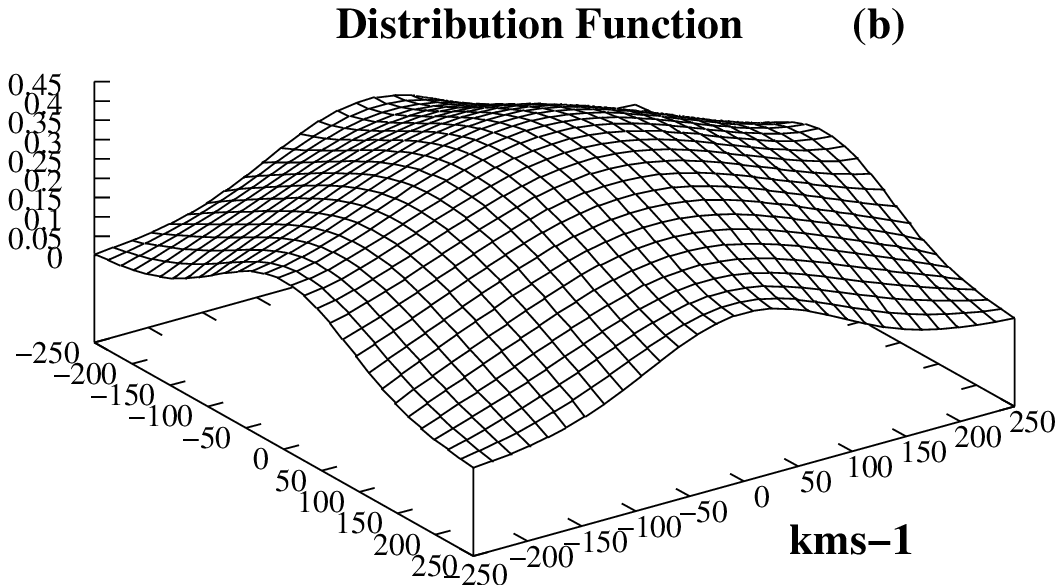}

\epsfig{width=8cm,height=6cm,file=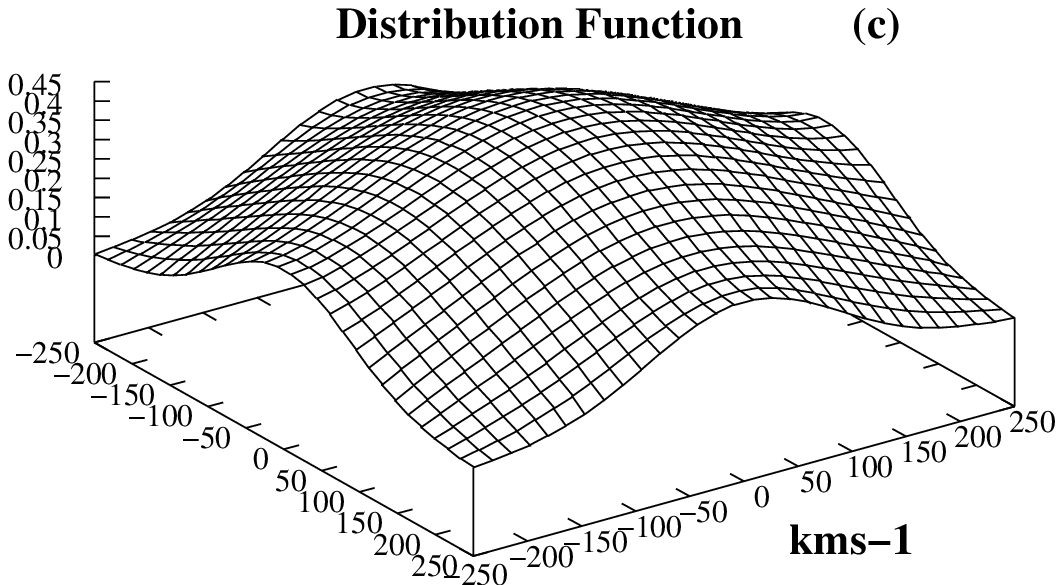}

\caption{Numerical solution of the velocity distribution of the
distribution function for a Kuzmin-Toomre disk type when $a=62.7$ kpc and
$\sigma_s= 40$ kpc. The velocity distribution functions
correspond to radius $r=$ 39.26 (a), 20.37 (b), 1.95 (c) kpc.}
\label{figperfvel}
\end{figure} 

\begin{figure}
\epsfig{width=8.5cm,height=6cm,file=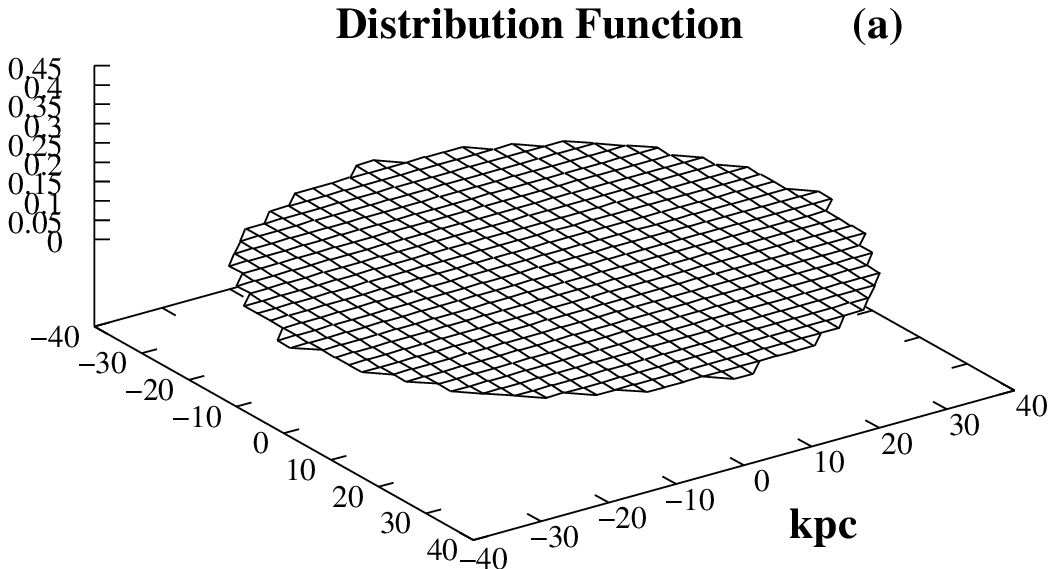}
 
\epsfig{width=8.5cm,height=6cm,file=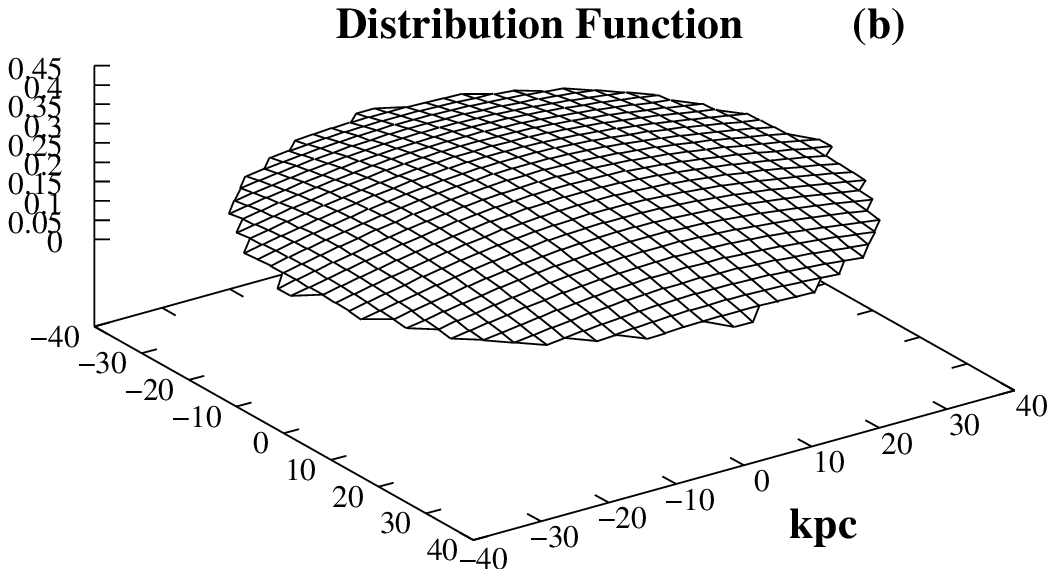}

\epsfig{width=8.5cm,height=6cm,file=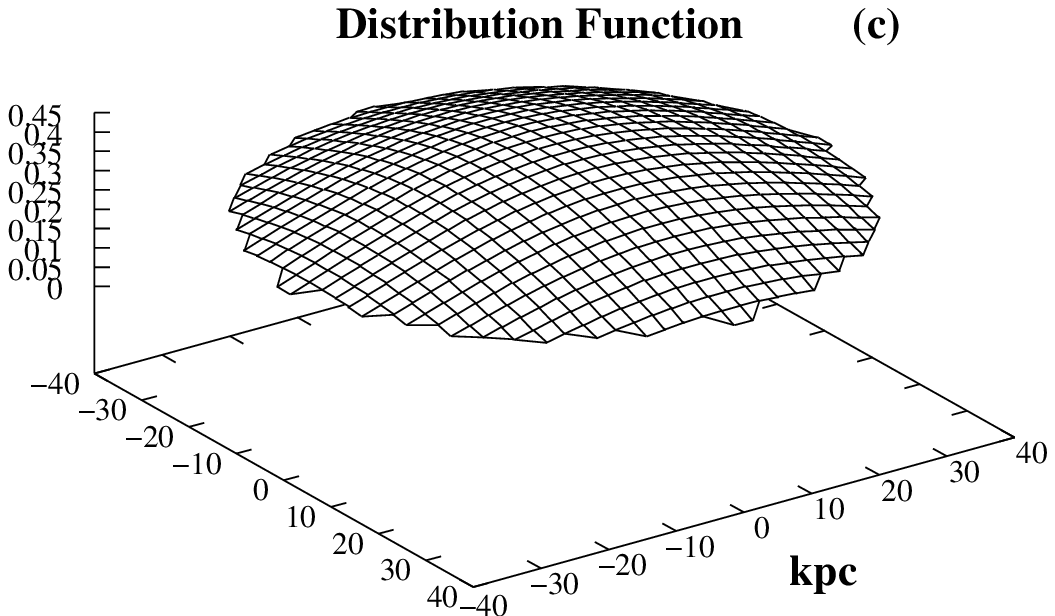}

\caption{Numerical solution of the spatial distribution of the
distribution function for a Kuzmin-Toomre disk type when $a=62.7$ kpc and
$\sigma_s =40$ kpc. The spatial distribution functions correspond to
velocities $v=\sqrt{v_x^2+v_y^2}=$ 341.6 (a), 207.3 (b), 27.3 (c) km
s$^{-1}$.}
\label{figperfpos}
\end{figure} 

For every internal node of the spatial part (full circles in Fig.
\ref{figgrid3}) we perform the integration of the distribution function
(\ref{mass}) obtaining in that way the surface mass density of the disk.
Note that the surface density is only known in these internal nodes.
This surface mass density can be plotted in a conventional three
dimensional plot ($x,y,\Sigma$). But, this makes the visual comparison
with the exact result $\Sigma_{TK}$ cumbersome. For ease in plotting and
comparison, we shall plot the surface mass density in a two dimensional
plot ($\{x,y\},\Sigma$). The nodes $\{x,y\}$ are chosen following the
direction of the arrows, starting from the left {\it start} arrow. Once
finished, we moved to the next right arrow and plot the $\{x,y\}$ nodes
of this arrow along the nodes of the previous arrow. We repeat this
procedure until we reach the {\it end} arrow. The nodes that are outside
the disk (empty circles) have surface mass density equal to zero and
will be omitted for the plot. We use this procedure for the numerical
solution as well as for the exact mass surface density solution. In the
later case, we take the exact values corresponding to the grid nodes
used in the numerical solution. We show the results in a two dimensional
figure, instead of a three dimensional one, because we think that the
value difference of the mass surface density between the grid points in
the numerical and exact solutions can be noted easily. The spatial
difference between arrows is the spatial discretization $\Delta x$ used
in the solution of the Fokker-Planck equation.

In the upper graph of Fig. \ref{density} we present the numerical result
of the mass surface density (\ref{mass}) with parameters $\sigma_s=35$
kpc, $\sigma_v=40$ km s$^{-1}$ and $a=40$ kpc (dotted line), compared
with the exact density function (solid line) obtain from the data of the
Kuzmin-Toomre density function (\ref{dentoom}). In this case $v_{esc}
\approx 460$ km s$^{-1}$ and the integration have been performed in the
intervals $v_x \in [-350,350]$ km s$^{-1}$ and $v_y \in [-350,350]$ km
s$^{-1}$. By varying the cut parameter $a$ and the spatial dispersion
$\sigma_s$, we can lower or raise the height of the ridges in the
computed curves. note that this modifications also leads to changes in
the escape velocity and integration intervals. In the lower graph of
Fig. \ref{density} we show the case with parameters $\sigma_s=40$ kpc,
$\sigma_v=40$ km s$^{-1}$ and $a=62.7$ kpc. For these parameters,
$v_{esc} \approx 370$, $v_x \in [-250,250]$ km s$^{-1}$ and $v_y \in
[-250,250]$ km s$^{-1}$. The maximum relative error between the computed
values and the exact values is in this case less than 1\%. This is a
good result because we are trying to match a real finite physical galaxy
to an infinite geometric potential that was developed without using any
observational data. In solving the Fokker-Planck equation we must take
into account that the final numerical result of the distribution
function has to be positive. Numerically, this means that all the values
of the distribution function at the grid nodes have to be positive.
Numerical simulations show us that this leads to restrictions in the
velocity dispersion and velocity intervals. For example, if we lowered
the values for the velocity dispersion to values near zero, the
distribution profile becomes very sharp and then some of the values of
the distribution function are negatives. The value of the velocity
dispersion depends on the galaxy model under consideration. In
particular, for the solution found above, we can not say that the
distribution function is unique because it depends on the relative error
allowed for the numerical solution. For example, if we allowed a
relative error of 1.5\%, the parameters $a$ and $\sigma_s$ can take
values between [62.4 kpc, 63 kpc] and [39.9 kpc, 40.1 kpc] respectively;
and $\sigma_v$ can not take values lower than 25 km s$^{-1}$ (because
some values of the distribution function are negatives) but might have
higher values. In general, for the gravitational potential
(\ref{pottoom}), we were able to model galaxies with positive
distribution functions for different parameters values, i.e.  galaxies
with total number of stars ranging from 10$^{10}$ to 10$^{12}$, typical
velocities from 100 km s$^{-1}$ to 300 km s$^{-1}$, velocity dispersions
from 20 km s$^{-1}$ to 80 km s$^{-1}$ and galaxies radius from 10 kpc to
80 kpc. So, in principle, we can find stationary solutions for many
typical galaxies.

Note that the potential we are using is from an infinite thin disk, and
that we have made a cut at an usual radius for galaxies. So, the results
we obtain are remarkable because we modeled a stationary galaxy with
physical parameters from a infinite-cut disk which do not necessarily
describes a real galaxy. Better agreement can be found if we enlarge the
value of the spatial dispersion above the fixed radius of the galaxy.
This behavior is equal for every size of the galaxy we choose, a
numerical indication that the galaxy is actually bigger than our chosen
radius. This is not surprising because the gravitational potential used
in the Fokker-Planck equation is originally from an infinite thin disk.

All the calculations have been done without actually knowing the
proportionality constant appearing in (\ref{boundary}). To determine the
proportionality constant, we take one point of the initial data for the
problem, the mass surface density $\Sigma_{TK}$ (\ref{dentoom}), and the
numerical mass surface solution at the same position. Later, the
numerical solution is scaled with this constant.

The numerical solution of the distribution function obtained in the
cases mentioned before can only be seen by a collection of three
dimensional figures. For example, if we want the velocity distribution
for the spatial part, we will have one figure for each pair ($x,y$). In
Fig.  \ref{figperfvel}, we show the velocity distribution function at
three different radius for a Kuzmin-Toomre thin disk with parameters
$\sigma_s=40$ kpc, $\sigma_v=40$ km s$^{-1}$ and $a=62.7$ kpc. From
these figures we note that, in order to have a stationary galaxy, the
velocity distribution function have to be centrally concentrated. In all
the cases, the population of stars with velocities that are near the
escape velocity of the system is almost equal to zero. Also, in Fig.
\ref{figperfpos} we present three spatial distribution functions for the
model of Fig. \ref{figperfvel}. We note from these figures that in
general the population of stars increases with lower velocities, and
that for each particular velocity the concentration of stars are bigger
near the center of the galaxy than in the edge of the galaxy. Note that
the distribution function can have other symmetries besides the axially
symmetry presented in the gravitational potential (\ref{pottoom})and
surface density (\ref{dentoom}) due to the form of equation
(\ref{sfokker}).

Note that the distribution function obtained from (\ref{sfokker}) has to
be positive and also, after the integration (\ref{mass}), the surface
density profile obtained has to match the initially given surface
density (\ref{dentoom}). These facts impose restrictions on the boundary
conditions and for that reason the boundary condition can not be set
arbitrarily. The boundary condition (\ref{boundary}) used for the
numerical calculation satisfies all the above requirements. We have
tried other boundary conditions by simple modifying the exponents of
equation (\ref{boundary}) and with other different functions without
obtaining good results.

\section{Conclusions}

We describe a method to find numerically a distribution function that
simultaneously satisfies the Fokker-Planck and Poisson equations
providing a self-consistent galaxy model for thin disks. This solution
have been found by solving the original Fokker-Planck equation avoiding
simplification using a previously developed code based on a direct
finite differences method.

As an example, we find distributions function of the family of
Kuzmin-Toomre thin disks. As far as we test, we are able to model many
self-consistent stationary galaxies using the most common physical
parameters known for galaxies. This is accomplished by adjusting the
parameters found in the gravitational potential, galaxy model and
boundary condition. In principle, we can find consistent astrophysical
model for other thin disk potentials. Once the solution is found, the
stability analysis of the models can be achieved performing
perturbations to the stationary solutions previously found and using the
same code presented in Ujevic \& Letelier (\cite{uje:let2}) and
performing a Crank-Nicolson discretization in time. We also found that
(\ref{boundary}) is an appropriate boundary condition for the case of a
stationary Kuzmin-Toomre stationary disk. In forthcoming articles this
condition will be checked for other thin disk models as well as three
dimensional disk models.

In this manuscript we presented the results for a case in which the
gravitational potential and the surface density is known a priori. This
was done only for didactic purposes because, in general, the method
presented here can be applied to more general problems in which we only
know the surface mass density of the disk. With this condition, the
gravitational potential can be obtained, numerically most of the time,
through the Poisson equation (\ref{poisson}), and then the distribution
function $f$ through the Fokker-Planck equation. Later with $f$, the
complete surface density is found via the integral (\ref{mass}). Then,
the problem is completely solved by matching the surface mass density
obtained via the integral (\ref{mass}) with the one found from the
Poisson equation.

For future applications, it will be very important to test this code and
boundary condition in other thin disk models and three dimensional
structures in order to calculate their distribution functions.  Three
dimensional models, can be obtained following the procedure of this
manuscript together with the code in Ujevic \& Letelier
(\cite{uje:let2}). In the last reference, a three dimensional
calculation of a test Fokker-Planck equation is presented. The
construction of a solution for Plummer and King models, thick disks, and
the stability of thin disks and other three dimensional structures are
currently under investigation by the authors. A stability analysis of
several thin disks using a perturbation on its energy-momentum tensor in
the context of general relativity can be found in Ujevic \& Letelier
(\cite{uje:let1}), another stability analysis of related structures
using Rayleigh criteria of stability can be found in Letelier
(\cite{let}).

\begin{acknowledgements}
M.U. and P.S.L. thanks FAPESP for financial support and P.S.L. also 
thanks CNPq.
\end{acknowledgements}

\appendix

\section{Projection into the $z-v_z$ plane} \label{deltas}

In this section, we develop some of the calculations that we encounter
in the substitution of the modified distribution function (\ref{delta})
into the stationary Fokker-Planck equation (\ref{sfokker}),

\begin{eqnarray}
&&\iint \frac{\partial}{\partial z} [ f \delta(z) \delta(v_z) ] dz dv_z = 
\int \left[ \frac{\partial f}{\partial z} \delta(z) + f \dot{\delta}(z) 
\right] dz \nonumber \\
&&\hspace{1.cm}= \left. \frac{\partial f}{\partial z} 
\right|_{z=0} - \left. \frac{\partial f}{\partial z} \right|_{z=0} = 0, 
\nonumber \\
&&\iint \frac{\partial}{\partial v_z} [ f D(\Delta v_z) \delta(z) 
\delta(v_z) ] dz dv_z \nonumber \\
&&\hspace{1.cm}= \int \left\{ \frac{\partial}{\partial v_z} [ 
f D(\Delta v_z) ] \delta(v_z) + f D(\Delta v_z) \dot{\delta}(v_z) \right\} 
dv_z \nonumber \\
&&\hspace{1.cm}= \left. \frac{\partial}{\partial v_z} [f D(\Delta v_z)] 
\right|_{v_z = 0} - \left. \frac{\partial}{\partial v_z} [f D(\Delta 
v_z)] \right|_{v_z = 0} = 0, \nonumber
\end{eqnarray}

\noindent where we used the well known Delta function properties:

$\int_{-\infty}^{+\infty} g(x) \delta(x-a) dx = g(a)$ and
$\int_{-\infty}^{+ \infty} g(x) \dot{\delta} (x-a) dx = - \dot{g}(a)$.

\noindent The remaining projections can be calculated in a similar way. 

\section{Two dimensional $LU$ decomposition} \label{lu}

\begin{table*}
\centering
\caption{\label{discret} Relations and nomenclature between the matrix
form and the one dimensional storage index at node $p$ of the thirteen
terms used in the discretization of the Fokker-Planck equation.}

\begin{tabular}{|cccc|}
\hline
Matrix Form & Abbreviation & Name & Position From Node $p$ \\
\hline
& \multicolumn{2}{c}{\rm \bf Basic Diagonals} &  \\
$f(i,j,k,l)$ & $P$ & point & $p$ \\
$f(i,j+1,k,l)$ & $N$ & north & $p+1$ \\
$f(i,j-1,k,l)$ & $S$ & south & $p-1$ \\
$f(i+1,j,k,l)$ & $E$ & east & $p+n_j$ \\
$f(i-1,j,k,l)$ & $W$ & west & $p-n_j$ \\
$f(i,j,k+1,l)$ & $T$ & top & $p+n_{ij}$ \\
$f(i,j,k-1,l)$ & $B$ & bottom & $p-n_{ij}$ \\
$f(i,j,k,l+1)$ & $U$ & up & $p+n_{ijk}$ \\
$f(i,j,k,l-1)$ & $D$ & down & $p-n_{ijk}$ \\
& \multicolumn{2}{c}{\rm \bf Mixed Diagonals} &  \\
$f(i,j,k+1,l+1)$ & $UT$ & uptop & $p+(n_{ijk} + n_{ij})$ \\
$f(i,j,k-1,l+1)$ & $UB$ & upbottom & $p+(n_{ijk} - n_{ij})$ \\
$f(i,j,k+1,l-1)$ & $DT$ & downtop & $p-(n_{ijk} - n_{ij})$ \\
$f(i,j,k-1,l-1)$ & $DB$ & downbottom & $p-(n_{ijk}+n_{ij})$ 
\\
\hline
\end{tabular}
\end{table*}

The $LU$ decomposition for the two dimensional version of the modified
Stone method presented in Ujevic \& Letelier (\cite{uje:let2}) is,

\begin{eqnarray}
&&L^p_{DB} = \frac{A^p_{DB}}{1+\alpha \left[ U^{p-(n_{ijk}+n_{ij})}_{N}
+U^{p-(n_{ijk}+n_{ij})}_{E} +U^{p-(n_{ijk}+n_{ij})}_{UB} \right]},
\nonumber \\
&&L^p_{D} = \frac{A^p_{D} - L^p_{DB} U^{p-(n_{ijk}+n_{ij})}_{T}}{1 +
\alpha \left[ U^{p-n_{ijk}}_{N} + U^{p-n_{ijk}}_{E} \right]}, \nonumber \\
&&L^p_{DT} =(A^p_{DT} - L^p_{D} U^{p-n_{ijk}}_{T})/\left\{1 + \alpha
\left[ U^{p-(n_{ijk}-n_{ij})}_{N} \right. \right. \nonumber \\
&&\hspace{1cm}\left. \left.+ U^{p-(n_{ijk}-n_{ij})}_{E} + 
U^{p-(n_{ijk}-n_{ij})}_{T} + 
U^{p-(n_{ijk}-n_{ij})}_{UT} \right]\right\}, \nonumber \\
&&L^p_{B} = \frac{A^p_{B} - L^p_{D} U^{p-n_{ijk}}_{UB} - L^p_{DB}
U^{p-(n_{ijk}+n_{ij})}_{U}}{1 +\alpha \left[ U^{p-n_{ij}}_{N} +
U^{p-n_{ij}}_{E} +U^{p-n_{ij}}_{UB} \right]}, \nonumber \\
&&L^p_{W} = \frac{A^p_{W}}{1+ \alpha \left[ U^{p-n_j}_{N} +
U^{p-n_j}_{T} + U^{p-n_j}_{UB} + U^{p-n_j}_{U} + U^{p-n_j}_{UT}
\right]}, \nonumber \\
&&L^p_{S} = \frac{A^p_{S}}{1 + \alpha \left[ U^{p-1}_{E} + U^{p-1}_{T}
+U^{p-1}_{UB} +U^{p-1}_{U} +U^{p-1}_{UT} \right]}, \nonumber \\
&&H_1 = \alpha \left[ L^p_{DB} U^{p-(n_{ijk}+n_{ij})}_{N} + L^p_{D}
U^{p-n_{ijk}}_{N} + L^p_{DT} U^{p-(n_{ijk}-n_{ij})}_{N} \right. \nonumber 
\\
&&\hspace{1cm}\left.+ L^p_{B} U^{p-n_{ij}}_{N} + L^p_{W} U^{p-n_j}_{N} 
\right], \nonumber \\
&&H_2 = \alpha \left[ L^p_{DB} U^{p-(n_{ijk}+n_{ij})}_{E} + L^p_{D}
U^{p-n_{ijk}}_{E} + L^p_{DT} U^{p-(n_{ijk}-n_{ij})}_{E} \right. \nonumber 
\\
&&\hspace{1cm}\left.+ L^p_{B} U^{p-n_{ij}}_{E} + L^p_{S} U^{p-1}_{E} 
\right], \nonumber \\
&&H_3 = \alpha \left[ L^p_{DT} U^{p-(n_{ijk}-n_{ij})}_{T} + L^p_{W}
U^{p-n_j}_{T} + L^p_{S} U^{p-1}_{T} \right], \nonumber \\
&&H_4 = \alpha \left[ L^p_{DB} U^{p-(n_{ijk}+n_{ij})}_{UB} + L^p_{B}
U^{p-n_{ij}}_{UB} + L^p_{W} U^{p-n_j}_{UB} + L^p_{S} U^{p-1}_{UB}
\right], \nonumber \\
&&H_5 = \alpha \left[ L^p_{W} U^{p-n_j}_{U} + L^p_{S} U^{p-1}_{U} \right],
\nonumber \\
&&H_6 = \alpha \left[ L^p_{DT} U^{p-(n_{ijk}-n_{ij})}_{UT} + L^p_{W}
U^{p-n_j}_{UT} + L^p_{S} U^{p-1}_{UT} \right], \nonumber \\
&&L^p_{P} = H_1 + H_2 +H_3 + H_4 + H_5 + H_6 - L^p_{S} U^{p-1}_{N} 
-L^p_{W} U^{p-n_j}_{E} \nonumber \\
&&\hspace{1.cm}- L^p_{B} U^{p-n_{ij}}_{T} - L^p_{DT} 
U^{p-(n_{ijk}-n_{ij})}_{UB} - L^p_{D} U^{p-n_{ijk}}_{U} \nonumber \\
&&\hspace{1.cm}- L^p_{DB} U^{p-(n_{ijk}+n_{ij})}_{UT}, \nonumber \\
&&U^p_{N} = \frac{A^p_{N} - H_1}{ L^p_{P}}, \nonumber \\
&&U^p_{E} = \frac{A^p_{E} - H_2}{ L^p_{P}}, \nonumber \\
&&U^p_{T} = \frac{A^p_{T} - H_3 - L^p_{DT} U^{p-(n_{ijk}-n_{ij})}_{U} -
L^p_{D} U^{p-n_{ijk}}_{UT}}{ L^p_{P}}, \nonumber \\
&&U^p_{UB} = \frac{A^p_{UB} - H_4 - L^p_{B} U^{p-n_{ij}}_{U}}{ L^p_{P}},
\nonumber \\
&&U^p_{U} = \frac{A^p_{U} - H_5 - L^p_{B} U^{p-n_{ij}}_{UT}}{ L^p_{P}},
\nonumber \\
&&U^p_{UT} = \frac{A^p_{UT} - H_6}{ L^p_{P}}. \nonumber
\end{eqnarray}

\noindent where $X^p_Y$ is the element $p$ of the diagonal $Y$ in
matrix $X$. These relations have to be calculated in the order
specified above. The notation used for the thirteen point molecule is
describe in Table \ref{discret}.

\end{document}